\documentclass[aps,prb,twocolumn,superscriptaddress,showpacs,10pt]{revtex4-2}
\usepackage{amsmath}
\usepackage[applemac]{inputenc}
\usepackage{color}
\usepackage{graphicx}
\usepackage{amssymb}
\usepackage{epsfig}
\usepackage{amsmath}
\usepackage{float}

\begin{document}

\title{Thermal Waves and Heat Transfer Efficiency Enhancement in Harmonically Modulated Turbulent Thermal Convection}

\author{P.~Urban} \author{P. Hanzelka} \author{T. Kr\'{a}l\'{i}k} \author{V. Musilov\'{a}}\affiliation{The Czech Academy of Sciences, Institute of Scientific Instruments, Kr\'{a}lovopolsk\'{a} 147, Brno, Czech Republic}
\author{L.~Skrbek} \affiliation{Faculty of Mathematics and Physics, Charles University, Ke Karlovu 3, 121 16, Prague, Czech Republic}

\begin{abstract}
We study turbulent Rayleigh-B\'{e}nard convection over four decades of Rayleigh 
numbers $4 \times 10^8 < {\rm{Ra}} < 2\times 10^{12}$, while harmonically modulating the temperatures of the plates of our cylindrical cell.  We probe the flow by temperature sensors placed in the cell interior and embedded in the highly conducting copper plates and detect thermal waves  propagating at modulation frequency in the bulk of the convective flow. We confirm the recent numerical prediction [PRL \textbf{125}, 154502 (2020)] of the significant enhancement of Nusselt number and report its dependence on the frequency and amplitude of the temperature modulation of plates. 

\today
\end{abstract}
\maketitle

Our everyday life is continuously affected by a plethora of natural and industrial turbulent flows, driven by time-dependent forcing. They include tidal ocean currents driven by the periodic gravitational  attraction of the Moon, the flows of the Earth's atmosphere due to the heat of the Sun, piston and combustion-generated flows in engines of cars and ships, or blood flow in our veins in tandem to the beating of our hearts. Still, periodically-driven turbulent flows have so far attracted much less attention than their statistically-steady counterparts, which is especially true for the class of buoyancy-driven flows. Here the paradigm is the turbulent Rayleigh-B\'{e}nard convection (RBC)~\cite{AhlersRevModPhys,ChillaSchumacher}, occurring within a layer of Oberbeck-Boussinesq (OB) fluid between two horizontal, ideally conducting and laterally infinite plates separated by the distance $L$ in a gravitational field (acceleration due to gravity $g$). Confined  
turbulent RBC, a very useful laboratory tool, belongs to the family of most frequently studied turbulent flows, however, studies of confined RBC driven by steady temperature difference $\Delta T_0$ with a weak superimposed sinusoidal temperature perturbation \cite{NiemelaSreeni,NiemelaPhysScripta,KuqaliNiemela} up to the case of fully periodic thermal drive~\cite{LohsePRL2020},
periodically forced \cite{Sterl} or kicked thermal turbulence \cite{JinXia} are scarce.

\begin{figure}[h]
\includegraphics[width=1\linewidth]{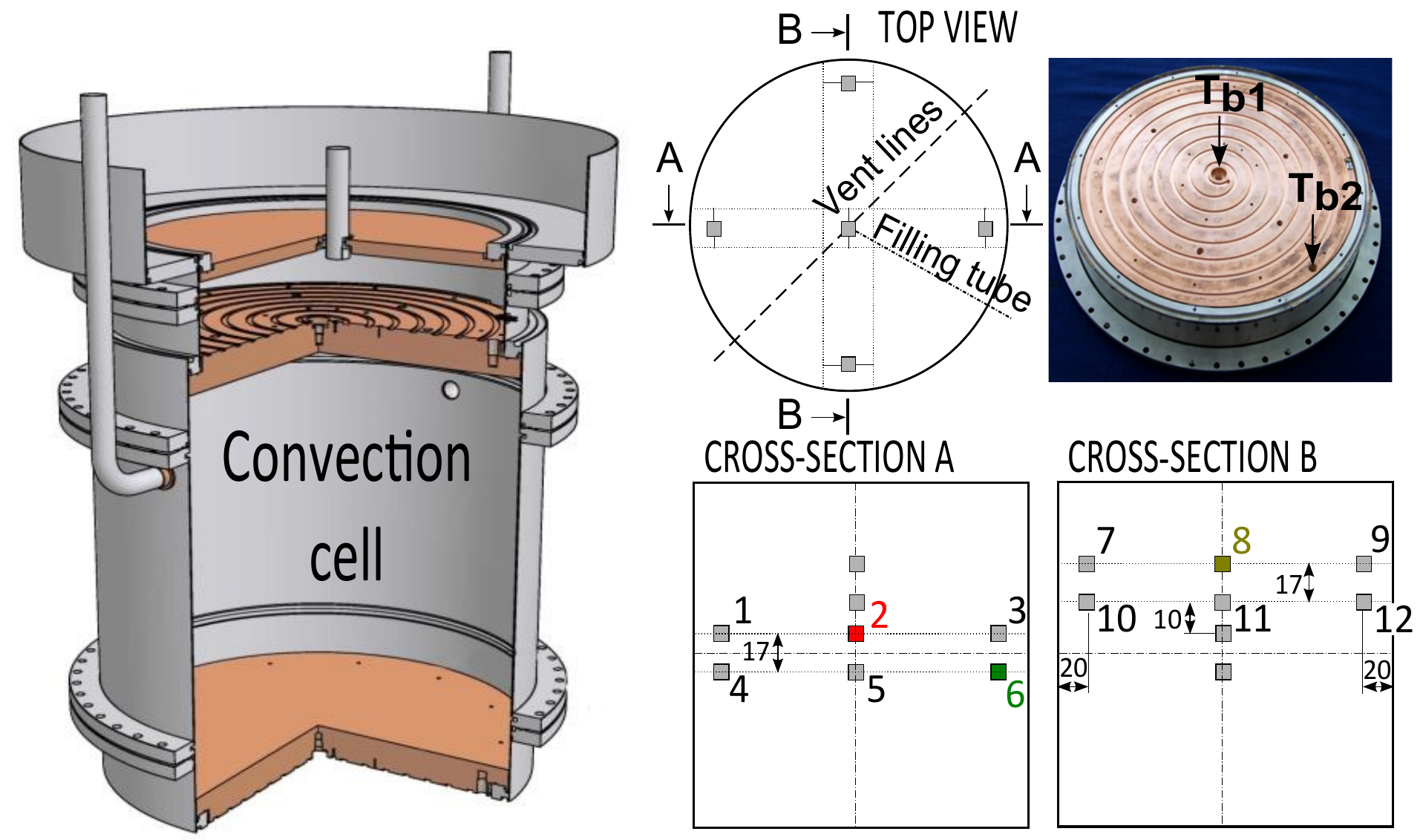}
\caption{Left: The $L=30$~cm tall aspect ratio $\mathit{\mathit{G}}=D/L=1$ RBC cell with 28 mm thick top and bottom plates $D=30$~cm in diameter made of thermally annealed copper of thermal conductivity $\lambda_p$=2210 W~m$^{-1}$K$^{-1}$ and thermal capacity $c_p$=0.144~Jkg$^{-1}$K$^{-1}$ at $T_{He}=(T_T + T_B)/2 \approx 5$~K, where $T_T$ and $T_B$ are typical temperatures of the top and bottom plates. From the top plate, most of the heat is removed via the heat exchange chamber to the liquid He vessel above it. The top plate temperature $T_T(t)$ is roughly set by pressure in the heat exchange chamber and more precisely tuned and modulated by the uniformly-distributed heater glued in the spiral grove on the upper side of the top plate. A similar heater delivers either steady or harmonically-modulated heat to the bottom plate. The temperature of the 
convective flow at locations as shown (distances in mm) is probed by small Ge sensors (numbered 1 ... 12) and that of the plates by
the finely calibrated Ge sensors $\rm {T}_{t1}, \rm {T}_{t2}, \rm {T}_{b1}$ and $\rm {T}_{b2}$ embedded in them; see the photograph in the top right, showing their positions and the spiral heater grove. }
\label{fig:BrnoCell}
\end{figure}

In this Letter, we present an experimental study of periodically modulated thermal convection in a cylindrical cell~\cite{BrnoCellRSI}, capable of reaching, under OB conditions, turbulent RBC flows over a wide range of Rayleigh numbers $10^8 \leq {\rm{Ra}} \leq 3 \times 10^{12}$ in a single experiment~\cite{BrnoPRL1,BrnoPRL2,OurNJP,OurJFM2017}. For harmonic modulation 
of the the mean temperature difference between the bottom and top plates, 
$\Delta T(t) = T_B(t)-T_T(t) = \Delta T_0 [1 + A_T \sin(2 \pi f_m t)]$, we map the relative depth $A_S(\mathbf{r})/A_T$ and the phase $\Phi(\mathbf{r})$ of the observed dimensionless temperature modulation $(A_S(\mathbf{r})/A_T) \sin(2 \pi f_m t+\Phi(\mathbf{r}))$ at various positions $\mathbf{r}$ in the cell. For $0 \leq A_T \leq 1$ we test the heat transfer efficiency by measuring the enhancement of Nusselt number, Nu, in dependence on $f_m$ and $A_T$ at either plate at various Ra, defined as 
${\rm{Ra}}= g \Delta T_0 L^3 \alpha/(\nu \kappa)$. Here $\alpha$, $\nu$ and $\kappa=\lambda/(\rho c_p)$ are, respectively, the thermal expansion, kinematic viscosity, thermal diffusivity, thermal conductivity, density and specific heat at constant pressure of our working fluid, cryogenic helium gas~\cite{Hepak}.

The RBC cell~\cite{BrnoCellRSI} is schematically shown in Fig.~\ref{fig:BrnoCell}.
Cryogenic conditions are essential, 
as the cell must be ``fast" in order to faithfully and uniformly follow temperature modulations imposed to its plates; for detailed discussion see \cite{EPLtoBe}. The harmonic modulation 
$\Delta T(t)$ 
is achieved by applying suitable time-dependent heat fluxes $q_B(t)$ and $q_T(t)$, using a home-made PID control scheme in such a way that the time-averaged temperature difference $\Delta T_0=\langle \Delta T(t) \rangle$ remains constant. The dimensionless efficiency of the heat transport, Nu, is defined via the mean value of the heat delivered to the bottom plate:  ${\rm{Nu}} = L\langle q_B(t) \rangle/(\lambda \Delta T_0)$.

One of motivations for this study was to verify and extend the work of Niemela and Sreenivasan~\cite{NiemelaSreeni}, 
performed in a similar RBC cell,  
who used a small Ge sensor at the half-height of the cell 
and measured the temperature signal which followed the harmonically modulated temperature of the bottom plate. For moderate ${\rm{Ra}}\approx 10^9$, the modulation 
of the vertically propagating thermal wave agreed with the form 
\begin{equation}
A_S(z,t)= A_T \left [ \exp (- \frac{z}{\delta_S}) \sin \left ( 2 \pi f_m t- \frac{z}{\delta_S} \right ) \right ]; 
\label{LohsePRL2020}
\end{equation}
where $\delta_S=(\pi f_m/\kappa_{eff})^{-1/2}$ denotes the Stokes layer thickness based on effective thermal diffusivity, characteristic of the turbulent working fluid, $\kappa_{eff}= {\rm{Nu}}\, \kappa$. For ${\rm{Ra}}\approx 10^{13}$ it was, however, not the case: significantly larger amplitude $A_S$ was observed. The authors explained their observations by suggesting 
a ``superconducting core" in the bulk, across which the thermal wave propagates without attenuation~\cite{NiemelaSreeni,NiemelaPhysScripta,KuqaliNiemela}. 

\begin{figure}[t]
\centering
\includegraphics[width=1\linewidth]{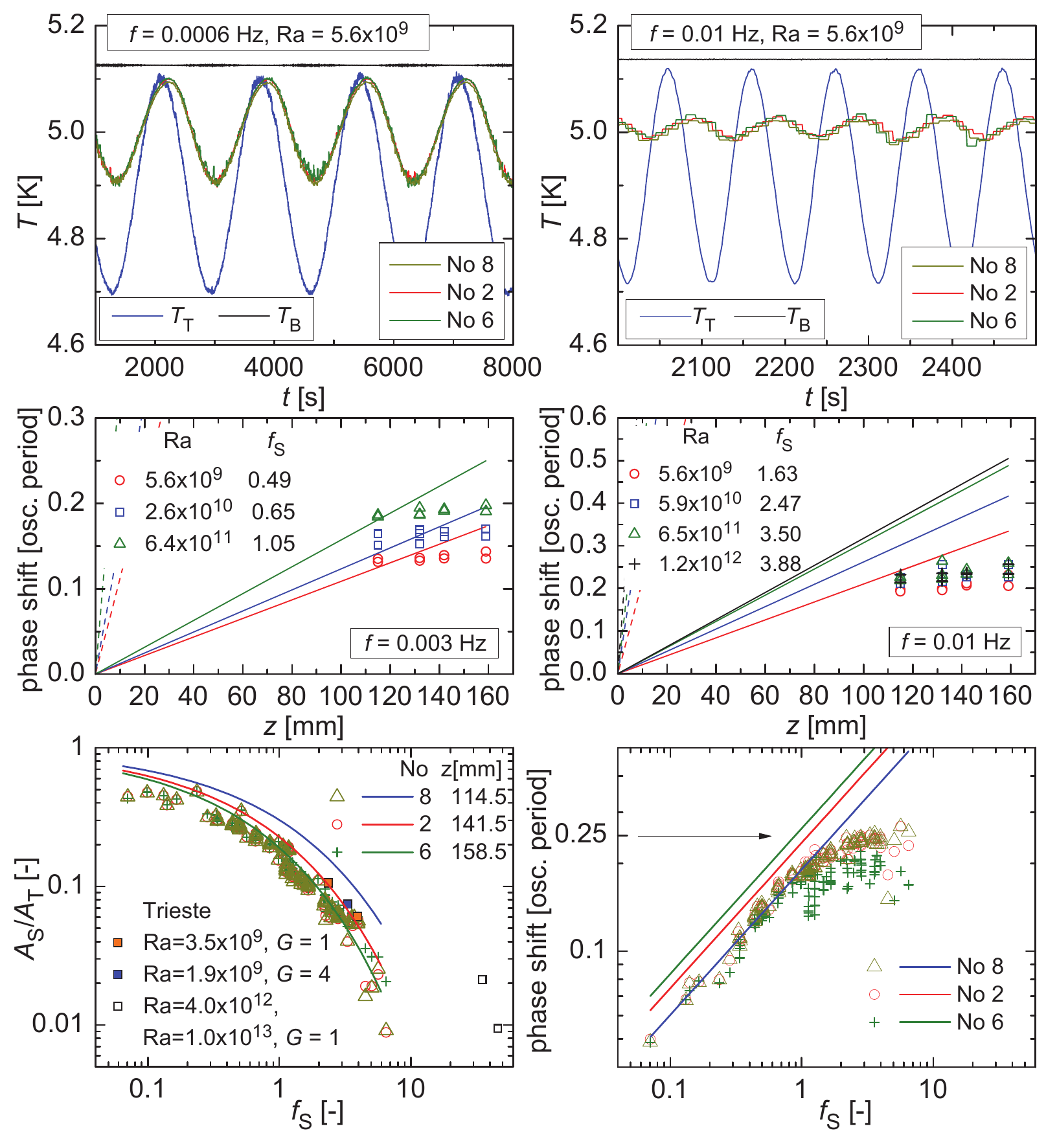}
\caption{Top panels: The records of the imposed harmonic modulation of the top plate $T_T(t)$ and constant $T_B(t)$ plotted together with the temperature records from sensors No 2, 6 and 8:  their readings are identical within experimental resolution. Middle panels: Examples of phase shifts of the thermal wave plotted versus the distance from the top plate for $f_S < \approx 1$ (left) and $f_S > 1$ (right) by the sensors in the bulk as indicated. The solid (dashed) straight lines are plots of phase shifts in Eq.~\ref{LohsePRL2020} with $\kappa_{eff}= {\rm{Nu}} \kappa$ ($\kappa_{eff}= \kappa$).
Bottom panels: The relative temperature amplitude $A_S/A_T$ (left) and the phase shift (right) measured by sensors as indicated at various $f_{m}$, $A_T$ and $\rm{Ra}$ as shown in Fig.~\ref{fig:Kvadrat} collapse when plotted versus 
$f_{S}$, defined by Eq.~\ref{eq:nondim}. The data points measured in Trieste~\cite{NiemelaSreeni,NiemelaPhysScripta} are included for comparison. 
The lines (blue, red and green - sensors No 8, 2 and 6, respectively) are plots of Eq.~\ref{LohsePRL2020} with $\kappa_{eff}= {\rm{Nu}}\, \kappa$. 
The phase shift saturates at $\pi/2$ as indicated by the arrow. 
}
\label{fig:Gap}
\end{figure}

Our cell allows for a more detailed investigation of modulated turbulent RBC flow, as we can measure the temperature signal in multiple positions, as detailed in Fig.~\ref{fig:BrnoCell}. 
As a result of harmonic temperature oscillations imposed on either plate, we indeed observe thermal waves propagating from the plates into the bulk of the turbulent flow. The situation is, however, more complex and cannot be described solely by Eq.~\ref{LohsePRL2020}, as we illustrate in Fig.~\ref{fig:Gap}. The readings from sensors No
8, 2 and 6, placed 
at increasing vertical distances from the top plate,
temperature of which is modulated, are
within experimental resolution identical in both amplitudes and phase shifts. This remarkable feature is observed for signals from any sensor in the bulk (see  Fig.~\ref{fig:BrnoCell}), over the entire range of investigated 
Ra \cite{sidewallSensor}. A similar but upside down outcome is observed when $T_B(t)$ is modulated instead of $T_T(t)$.

We therefore extend the model of propagating thermal waves leading to Eq.~\ref{LohsePRL2020}, by considering the vertical structure of the high Ra turbulent convective flow, namely the existence of thermal and kinetic boundary layers (BL).
For cryogenic He gas of ${\rm{Pr}} \approx 1$ the situation is simpler in that both BL are of similar thicknesses: $\ell_{BL} \approx L/(2{\rm{Nu}})$. In accord with the dimensional reasoning of Priestley \cite{Priestley} and marginal stability arguments of Malkus \cite{Malkus}, the high Ra turbulent convective flow consists of (almost) isothermal bulk and two laminar BLs adjacent to plates, over which most of the temperature difference drops. With increasing Ra, $\ell_{BL}$ decreases and the temperature profile near the plates becomes steeper. Within the BL, the application of Eq.~\ref{LohsePRL2020} is justified, with $\kappa_{eff} \approx \kappa$. The initial slope of lines (plots of Eq.~\ref{LohsePRL2020}) shown in the middle panels of Fig.~\ref{fig:Gap} therefore must be $\sqrt{Nu}$ times steeper (dashed lines). In order to appreciably affect the bulk, the thermal wave has to pass the BL, where its amplitude decreases exponentially with characteristic distance $\delta_S^{BL}= (\pi f_m/\kappa)^{-1/2}$. As soon as the thermal wave passes the BL, it is expected to propagate in the bulk according to Eq.~\ref{LohsePRL2020}, where $\kappa_{eff}= {\rm{Nu}}_{loc}\, \kappa$, where ${\rm{Nu}}_{loc}$ is the local Nusselt number. Experimental observations just described suggest that ${\rm{Nu}}_{loc}$ in the bulk must be very large (consistent with the suggestion of a ``superconducting core" \cite{NiemelaSreeni}), in accord with simulations of high Ra RBC with periodic boundary conditions (i.e., effectivelly excluding BLs)~\cite{LohseToschi}.

These considerations allow to find a characteristic frequency, emerging from the condition of equal Stokes layer and BL thicknesses: $f_{onset}=4\kappa {\rm{Nu}}^2/(\pi L^2)$, yielding for our cell $(2<f_{onset}<7)$~Hz over the investigated range of Ra. A significant attenuation of temperature wave within the BL therefore should occur at these frequencies, which are much higher than those used in our experiment: $(0.0060 \le f_{m} \le 0.1)$~Hz. Consequently, anticipating only a weak effect of BL on the temperature wave, we confront our experimental results with Eq.~\ref{LohsePRL2020}. For each $\delta_S$,
we introduce a nondimensional scaling frequency $f_S$. Assuming validity of Eq.~\ref{LohsePRL2020}, the data on a phase shift or amplitude attenuation of temperature wave measured at a distance $z$ from the plate should collapse on a universal dependence on $f_S$. Specifically, we introduce $f_S$ via the ratio of the cell double-height $2L$ and the wavelength $\lambda_S=2\pi\delta_S$ of the wave in Eq.~\ref{LohsePRL2020}; i.e., as  $f_S = (2L/\lambda_S)^2= (L/\pi\delta_S)^2 = 4 L^2 f_m /(\pi \kappa \rm{Nu})$. Further, over the investigated range of Ra, the experimentally established scaling reads ${\rm{Nu}}/{\rm{Ra}}^{1/3}= \xi({\rm{Ra}})$, where $\xi({\rm{Ra}})$ is a numerical factor weakly decreasing with Ra, being about 0.07 at ${\rm{Ra}}= 10^8$, 0.06 at ${\rm{Ra}}= 10^{10}$ and about 0.055 at ${\rm{Ra}} > 10^{11}$ \cite{BrnoPRL1,BrnoCellRSI}, see Fig~\ref{fig:NuRa}.
By using relations for free fall time, $\tau_{ff} = \sqrt{L/(\alpha g \Delta T_0)} = ({\rm{RaPr}})^{-1/2}L^2/\kappa$ and ${\rm{Nu}}=\xi({\rm{Ra}}) {\rm{Ra}}^{1/3}$, we arrive at:
\begin{equation}
f_S=\frac{1}{\pi \xi({\rm{Ra}})} \hat{f} {\rm{Ra}}^{1/6} {\rm{Pr}}^{1/2}\,,
\label{eq:nondim}
\end{equation}                                                                  
\noindent
with nondimensional $\hat{f}=f_m \tau_{ff}$.

\begin{figure}[t]
\centering
\includegraphics[width=1\linewidth]{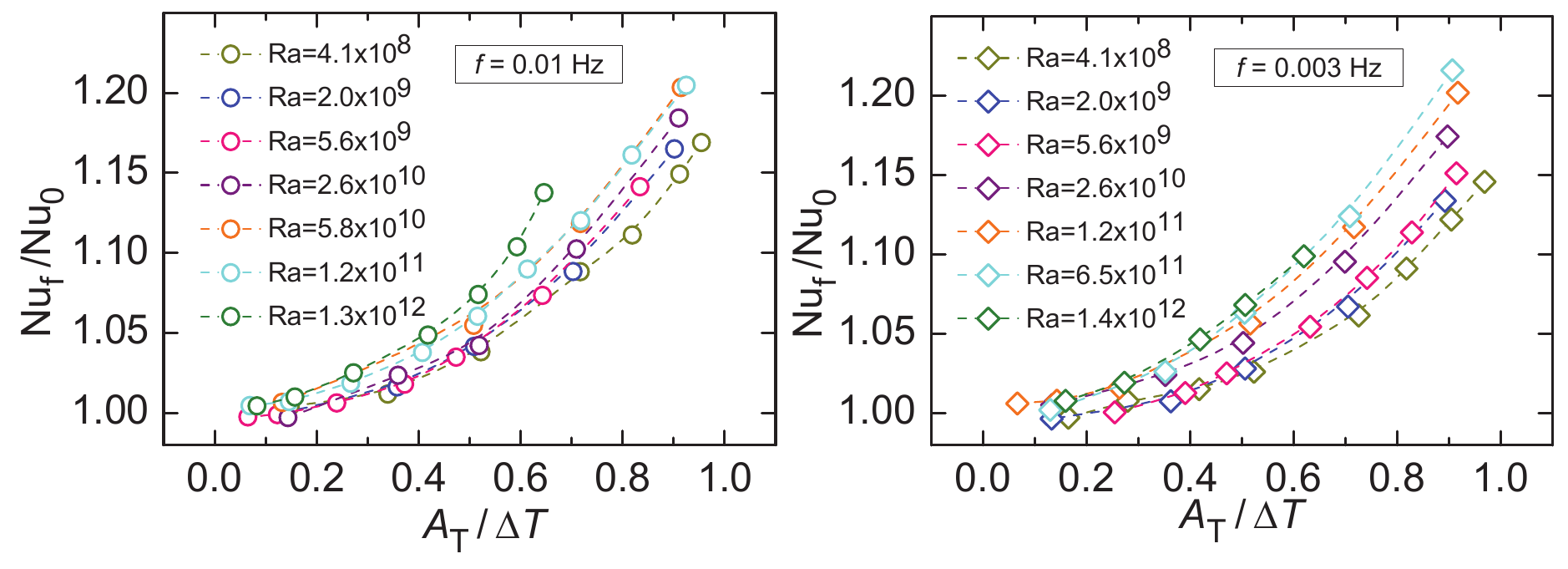}
\caption{Enhancement of ${\rm{Nu}}_f/{\rm{Nu}}_0$ plotted versus non-dimensional amplitude of temperature modulation $A_T$ applied at frequencies $f_m= 0.01$~Hz (left) and 0.003~Hz  (right) to the top plate, measured at $\rm{Ra}$ as indicated. The dashed lines represent the best power law fits.
}
 \label{fig:Kvadrat}
\end{figure}

The bottom panels of Fig.~\ref{fig:Gap} show the attenuated amplitudes and phase shifts measured by sensors in the bulk at various $\rm{Ra}$, $f_m$ and $A_T$, versus $f_S$. 
Plotted this way, the data points measured over three decades of $f_m$ and four decades of Ra, collapse on a single frequency dependence. In parallel, up to the $f_S \approx 1$ (remember the choice of factor 2 in the definition of $f_S$) the data follow qualitatively Eq.~\ref{LohsePRL2020} with Stokes thickness characterizing the thermal wave in the bulk. According to Eq.~\ref{LohsePRL2020} the wavelength equals $2L$ at $f_S=1$ and its phase at the mid-height of the cell reaches $\pi/2$. Contrary to Eq.~\ref{LohsePRL2020}, the measured phase shift saturates to this value for $f_S>1$ (as marked by an arrow in the bottom right panel of Fig.~\ref{fig:Gap}). As already mentioned, another disagreement is that the differences between individual sensor's readings do not depend 
(at least for the sensors on the cell axis~\cite{sidewallSensor}) on their position. Specifically, mutual differences in amplitudes and phases are at least by an order of magnitude smaller than those predicted by Eq.~\ref{LohsePRL2020}. On the whole, our results on detection and propagation of the thermal wave in modulated turbulent convective flow agree with measurements of Refs.~\cite{NiemelaSreeni,NiemelaPhysScripta} at the middle-height of their $\mathit{\mathit{G}}=1$ and $\mathit{\mathit{G}}=4$ cells for moderate $\rm{Ra}$ of order $10^9$; these data can be semi-quantitatively described by Eq.~\ref{LohsePRL2020}. On the other hand, this model is too crude for general description of the thermal wave in RBC, especially at higher $\rm{Ra}$, and cannot accurately describe all observed details.

\begin{figure}[t]
\centering
\includegraphics[width=1\linewidth]{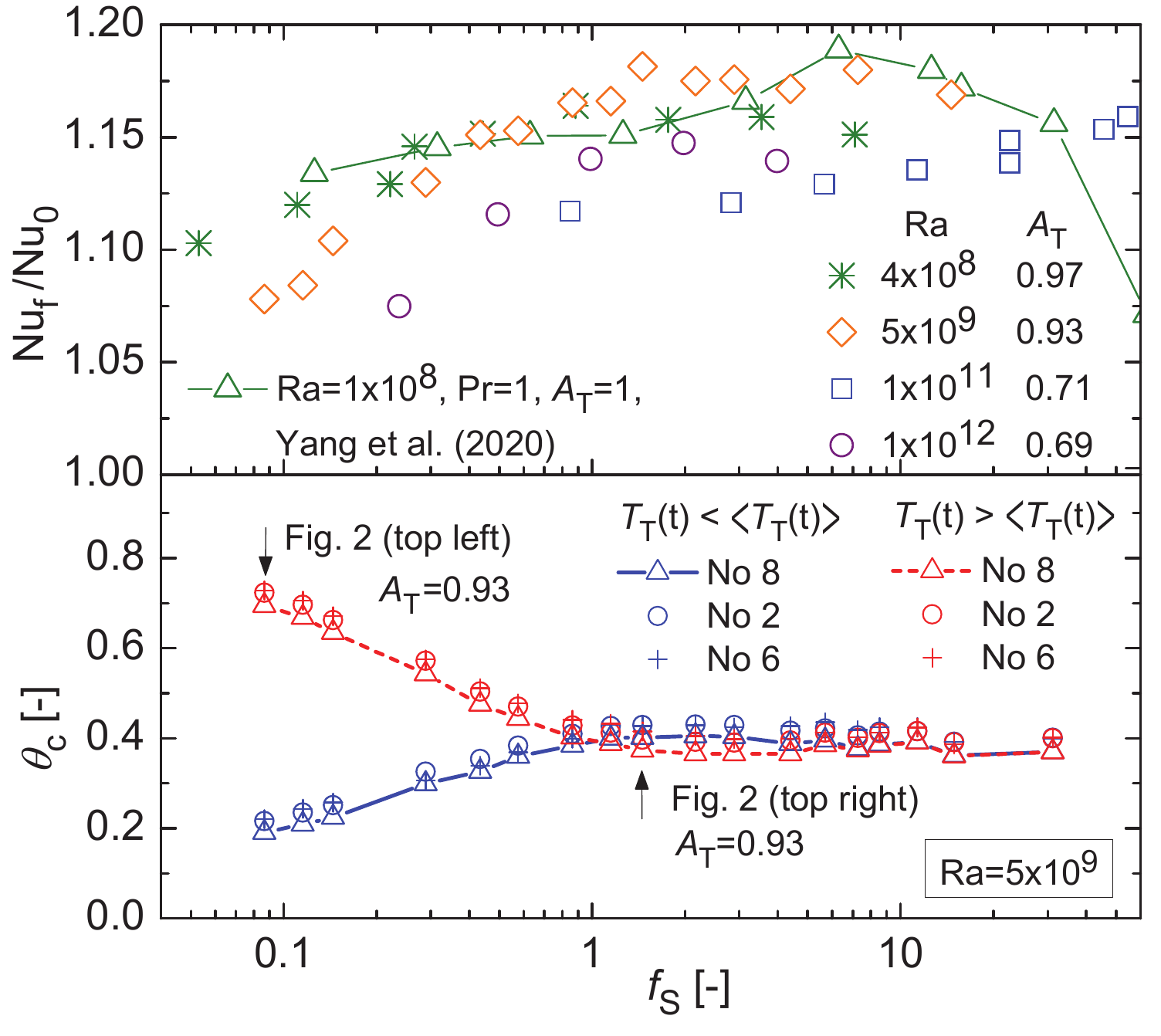}
\caption{Top: The relative enhancement ${\rm{Nu}}_f/{\rm{Nu}}_0$ versus 
$f_{s}$, measured for various relative temperature amplitudes $A_T$ applied to the top plate at ${\rm{Ra}}$ as indicated. The simulation data of Yang \textit{et al.}~\cite{LohsePRL2020} are plotted  
for comparison.
Bottom: Phase-averaged bulk dimensionless temperature $\theta_c$ measured by sensors 
No 2, 6 and 8. The red (blue) symbols represent the bulk temperature averaged over the half period of modulation when $T_T(t)$ is maximal (minimal). Frequencies $f_S$ used in the top panels of Fig.~\ref{fig:Gap} are indicated by arrows. 
} 
\label{fig:Freq}
\end{figure}

Another motivation for our study was that while the experiments \cite{NiemelaSreeni,NiemelaPhysScripta} do not report any enhancement of the heat transfer efficiency by harmonically modulated turbulent RBC flow, another experimental study, of kicked thermal turbulence, does. Specifically, Jin and Xia \cite{JinXia} report a resonant Nu enhancement of up to 7\%, found when the input pulse period was synchronized with half of the large-scale circulation (LSC) turnover time $\tau_{LSC}=1/f_{LSC}$. Moreover, the recent numerical 2D study (backed up by a 3D study for ${\rm{Ra}}=10^8$) of periodically modulated thermal convection by Yang \textit{et al.}~\cite{LohsePRL2020} predicts, for moderate Ra in the range $10^7 \leq {\rm{Ra}} \leq 10^9$, an appreciable enhancement of Nu, of up to about 25\%, occurring over a wide range of modulation frequencies lower or of order $f_{LSC}$. We therefore decided to perform experiments aiming at bridging these apparently contradictory findings.

Fig.~\ref{fig:Kvadrat} shows examples of the relative enhancement of the heat transfer efficiency, ${\rm{Nu}}_f/{\rm{Nu}}_0$, measured over four decades of Ra, plotted as a function of dimensionless harmonic temperature modulation $0 \leq A_T \leq 1$, applied at frequencies 0.003~Hz and 0.01~Hz to the top plate. The enhancement grows slightly steeper with increasing Ra, roughly with $A_T^2$; this perhaps explains why no enhancement was noticed for rather low values of $A_T \leq 0.22$ used in \cite{NiemelaSreeni,NiemelaPhysScripta}. We observe similar behavior of ${\rm{Nu}}_f/{\rm{Nu}}_0$ when the same temperature modulation is applied to either the top or bottom plate. From an experimental point of view, however, it is easier to apply the temperature modulation to the top plate. Delivering the additional heat flux needed for fast enough harmonic heating of either plate by using our home-made PID control is relatively easy, as our plates are ``thermally fast"~\cite{EPLtoBe}. That said, the bottom plate is cooled only via the convective heat flow in the cell, while the thermal connection of the top plate to the liquid helium bath above it via the heat exchange chamber is more efficient and can be set as needed \cite{BrnoCellRSI}. As a result, the top plate faithfully follows a harmonic temperature modulation up to higher $f_m$ \cite{BottomPlate}.

\begin{figure}[t]
\centering
\includegraphics[width=1\linewidth]{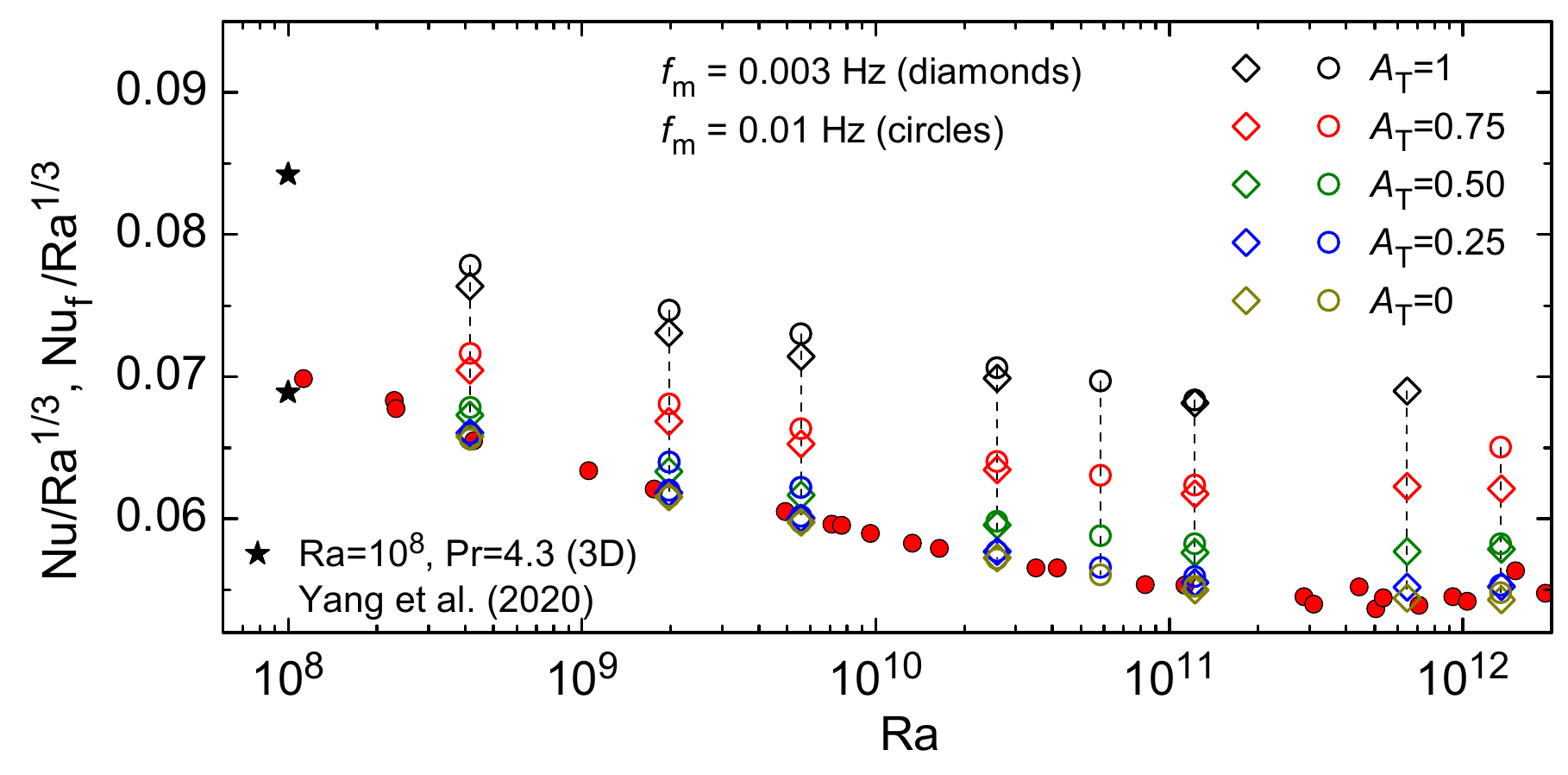}
\caption{
Relative enhancement of heat transfer efficiency, ${\rm{Nu}}_f/{\rm{Ra}}^{1/3}$, in harmonically-modulated turbulent convection. The data points for modulation amplitudes $A_T=0; 0.25; 0.5; 0.75; 1$ and frequencies $f_m$ as indicated are evaluated based on power law fits of the measured data as shown in Fig.~\ref{fig:Kvadrat}. The corresponding steady-state RBC data for constant heat flux to the bottom plate  and constant $T_T$ (red circles) \cite{OurJFM2017} and 3D
simulation data [6] are shown for comparison.
} 
\label{fig:NuRa}
\end{figure}

The top panel of 
Fig.~\ref{fig:Freq} displays ${\rm{Nu}}_f/{\rm{Nu}}_0$ as a function of $f_S$. In contrast with the resonant enhancement synchronized with half of the LSC turnover time $\tau_{LSC}=1/f_{LSC}$ found for the kicked turbulent convection \cite{JinXia}, the enhancement of the heat transfer efficiency in harmonically modulated turbulent convective flow occurs over several decades of $f_m$, in fair agreement with numerical prediction of Yang \textit{et al.}~\cite{LohsePRL2020} for ${\rm{Ra}}=10^8$ and ${\rm{Pr}}=1$.

The bottom panel of Fig.~\ref{fig:Freq} shows the typical phase-averaged dimensionless bulk temperature $\theta_c$. 
For $A_T$ and $\rm{Ra}$ as indicated it is defined via the bulk temperature $T_S(t)$ measured by sensors No 2, 6 and 8 as:
$\theta_c=(\langle T_S(t) \rangle_{hp}- \langle T_T(t) \rangle)/(T_B - \langle T_T(t) \rangle)$,
where $\langle ... \rangle_{hp}$ denotes
averaging over the half period of modulation when $T_T(t)$ is maximal (minimal). All the data series of this kind, measured at various $\rm{Ra}$ and $A_T$ and phase-averaged over the top and bottom half-periods of the modulation, split around the maximum of the observed ${\rm{Nu}}_f/{\rm{Nu}}_0$ (where $f_S \approx 1$ and the phase shift saturates to $\pi/2$) and display ``forks", in qualitative agreement with numerical prediction -- see Fig.~2f and relevant discussion in Ref.~\cite{LohsePRL2020}. Again, for cryogenic helium gas of ${\rm{Pr}} \approx 1$ the situation is simpler thanks to the fact that thermal and kinetic BLs are of similar thicknesses.

Fig.~\ref{fig:NuRa} illustrates the main result of our study: significant enhancement of heat transfer efficiency in harmonically-modulated convection ($f_m$ and $A_T$ applied to the top plate as indicated) in comparison with the steady-state RBC. 
We emphasize that all data were measured under identical experimental conditions and the very
same basic corrections~\cite{OurJFM2017}, due to adiabatic gradient and parasitic
heat leak, have been applied.  A key result is that the enhancement of compensated ${\rm{Nu}}$ is robust, occurring over broad ranges of governing flow parameters such as ${\rm{Ra}}$, $f_m$ and $A_T$. 

We believe that our results will stimulate further work  on the class of  periodic buoyancy-driven turbulent convective flows. This would lead to deeper understanding of important natural flows such as Sun-driven planetary weather formation, especially on Earth, including ocean and tidal flows or ice melting with direct relevance to global warming. There is also a practical aspect of our results -- their potential use in the design of more effective heat exchangers for various technical applications.

The authors thank M. Macek and K.R. Sreenivasan for stimulating discussions. This research is funded by the Czech Science Foundation under GA\v{C}R 20-00918S.


\begin{thebibliography}{}
\bibitem{sidewallSensor}
In some cases, we observed that the sensor No 6, placed close to the sidewall, shows even slightly smaller phase shift than sensors No 2 and 8, although its distance from the top plate is larger. We believe that this could occur due to the advection by the LSC flow.
\bibitem{BottomPlate}
The case when the total temperature modulation is applied at any ratio in the antiphase to both plates simultaneously is different and will be discussed separately.
\end{thebibliography}
\end{document}